\documentclass[twocolumn, amsmath, amssymb, longbibliography,superscriptaddress ]{revtex4-1}
\pdfoutput=1

\usepackage{graphicx}
\usepackage{dcolumn}
\usepackage{bm}
\usepackage{bbm}
\usepackage{color}
\usepackage{url}
\usepackage{multirow} 
\usepackage[normalem]{ulem}
\usepackage{xcolor}

\usepackage{epsfig}
\usepackage{color}
\usepackage{amsmath}
\usepackage{amssymb}
\usepackage{longtable}
\usepackage{dcolumn}
\usepackage{bm}
\usepackage{ulem}
\usepackage{float}
\usepackage{mathtools}
\usepackage{xparse}
\NewDocumentCommand{\ceil}{s O{} m}{%
  \IfBooleanTF{#1} 
    {\left\lceil#3\right\rceil} 
    {#2\lceil#3#2\rceil} 
}

\usepackage{algorithm}
\usepackage{algcompatible}

\newcommand{\be}{\begin{equation}}
\newcommand{\ee}{\end{equation}}
\newcommand{\bd}{\begin{displaymath}}
\newcommand{\ed}{\end{displaymath}}
\newcommand{\BE}{\begin{eqnarray}}
\newcommand{\EE}{\end{eqnarray}}

\newcommand{\bs}{\ensuremath{\mathbf{s}}}
\newcommand{\bu}{\ensuremath{\mathbf{u}}}
\newcommand{\bv}{\ensuremath{\mathbf{v}}}

\newcommand{\bz}{\ensuremath{\mathbf{z}}}

\newcommand{\fE}{\mathbb{E}}

\newcommand{\mcS}{\mathcal{S}}
\newcommand{\mcR}{\mathcal{R}}
\newcommand{\mcG}{\mathcal{G}}
\newcommand{\mcZ}{\mathcal{Z}}
\newcommand{\mcL}{\mathcal{L}}
\newcommand{\mcF}{\mathcal{F}}

\DeclareMathOperator{\sign}{sign}

\begin{document}

\preprint{}

\title{Quantum-assisted Helmholtz machines: A quantum-classical deep learning framework for industrial datasets in near-term devices}

\author{Marcello Benedetti}
\affiliation{Quantum Artificial Intelligence Lab., NASA Ames Research Center, Moffett Field, CA 94035, USA}
\affiliation{USRA Research Institute for Advanced Computer Science, Mountain View, CA 94043, USA}
\affiliation{Department of Computer Science, University College London, WC1E 6BT London, UK}
\affiliation{Cambridge Quantum Computing Limited, CB2 1UB Cambridge, UK}

\author{John Realpe-G\'omez}
\affiliation{Quantum Artificial Intelligence Lab., NASA Ames Research Center, Moffett Field, CA 94035, USA}
\affiliation{SGT Inc., Greenbelt, MD 20770, USA}
\affiliation{Instituto de Matem\'aticas Aplicadas, Universidad de Cartagena, Bol\'ivar 130001, Colombia}

\author{Alejandro Perdomo-Ortiz}
\email{Correspondence: alejandro.perdomoortiz@nasa.gov}
\affiliation{Quantum Artificial Intelligence Lab., NASA Ames Research Center, Moffett Field, CA 94035, USA}
\affiliation{USRA Research Institute for Advanced Computer Science, Mountain View, CA 94043, USA}
\affiliation{Department of Computer Science, University College London, WC1E 6BT London, UK}
\affiliation{Cambridge Quantum Computing Limited, CB2 1UB Cambridge, UK}
\affiliation{Qubitera, LLC., Mountain View, CA 94041, USA}

\begin{abstract}
Machine learning has been presented as one of the key applications for near-term quantum technologies, given its high commercial value and wide range of applicability. In this work, we introduce the \textit{quantum-assisted Helmholtz machine:} a hybrid quantum-classical framework with the potential of tackling high-dimensional real-world machine learning datasets on continuous variables. 
Instead of using quantum computers only to assist deep learning, as previous approaches have suggested, we use deep learning to extract a low-dimensional binary representation of data, suitable for processing on relatively small quantum computers. Then, the quantum hardware and deep learning architecture work together to train an unsupervised generative model. We demonstrate this concept using 1644 quantum bits of a \mbox{D-Wave 2000Q} quantum device to model a sub-sampled version of the MNIST handwritten digit dataset with 16 $\times$ 16 continuous valued pixels.
Although we illustrate this concept on a quantum annealer, adaptations to other quantum platforms, such as ion-trap technologies or superconducting gate-model architectures, could be explored within this flexible framework.
\end{abstract}

\maketitle

\section{Introduction}~\label{s:introduction}

There has been much interest in quantum algorithms for enhancing deep learning and other machine learning (ML) algorithms~\cite{neven2009nips, bian2010ising, Denil-2011, wiebe2012quantum, Pudenz-QIP-2013, Lloyd-arXiv-2013, Rebentrost-PRL-2014, wang2017quantum, 2015arXiv151203929Z, Lloyd-NatPhys-2014, schuld2016prediction, Wiebe-arXiv-2015, Aaronson-2015, Benedetti-2016, Benedetti2017, Adachi-arXiv-2015, chancellor2016maximum, Potok2017, Amin-arXiv-2016, kieferova2016tomography, kerenidis2016quantum, wittek2017quantum, Schuld-QML-2015, Romero2017, adcock2015advances, biamonte2016quantum, alvarez2016quantum, lamata2017basic, schuld2017quantum, 2017arXiv170708561C, PerdomoOrtiz2017,Benedetti2018,Farhi2018}. In this article, instead, we argue that deep learning and quantum devices can help each other to achieve hard tasks such as generative modeling. The resulting quantum-assisted ML (QAML) approach is much more suitable for implementation in near-term quantum hardware and can be used in real applications as well. Indeed, previous work has shown experimental evidence of the ability of quantum annealers to perform useful and realistic ML tasks, such as implementing generative models of small binarized datasets~\cite{Benedetti-2016,Adachi-arXiv-2015,chancellor2016maximum,Benedetti2017,Potok2017}. A natural extension is to develop techniques to handle large datasets---where variables could be discrete, continuous, or more general objects---and to include latent variables to increase the modeling capacity of the quantum-assisted architectures. Clearly, this would open up the possibility to use QAML in real-world domains and benchmark it against extensively studied classical approaches. This extension is the focus of this work.

The interest in generative models stems from their generality. Deep generative models with many layers of hidden stochastic variables have the ability to learn multi-modal distributions over high-dimensional datasets~\cite{bengio2009learning}. Each additional layer provides an increasingly abstract representation of the data and improves the generalization capability of the model~\cite{hinton2006fast}. Furthermore, generative models apply to unlabeled data, which accounts for most of the public data in the Internet and most of the private data in a company. Often, the price to pay for using a generative model is the intractability of inference, training, and model selection. Generative models are trained in an unsupervised fashion, relying on variational approximations and computationally expensive Markov Chain Monte Carlo (MCMC) sampling. This is where we think quantum computation can have a significant impact. Under the hypothesis that quantum computers allow more efficient sampling, we can run the expensive subroutine on quantum hardware. This would also enable us to exploit the non-trivial graph topologies in quantum hardware to implement complex networks, usually avoided in favor of restricted ones (e.g. bipartite graphs are favored in classical neural networks for convenience).

Quantum information does not have to be encoded into binary observables (qubits), it could also be encoded into continuous observables~\cite{lloyd1999quantum}. Some researchers have followed the latter direction~\cite{lau2017quantum, 2017arXiv170700360D}. However, most available quantum computers do work with qubits, nicely resembling the world of classical computation. Yet, datasets commonly found in industrial applications have a large number of variables that are not binary. For instance, datasets of images with millions of pixels which can be in gray scale, with $256$ intensities per pixel, or in color, represented by $3$-dimensional vectors. We refer to this kind of datasets as complex ML datasets. A naive binarization of the data will quickly consume the qubits of any device with ~100-1000 qubits. Several QAML algorithms~\cite{wiebe2012quantum, Rebentrost-PRL-2014, schuld2016prediction} rely on amplitude encoding instead, a technique where continuous data is stored in the amplitudes of a quantum state. This provides an exponentially efficient representation upon which one could perform linear algebra operations. Unfortunately, it is not clear how to prepare arbitrary states of this kind in near-term quantum computers. Reading out all the amplitudes of an output vector, if required by the application, might kill or significantly hamper any speedup~\cite{Aaronson-2015}. 

Here, we suggest using a quantum device to model an abstract representation of the data, that is, the deepest layers of a deep learning architecture. The number of hidden variables in the deepest layers of a network can indeed be much smaller than the number of visible variables, which is ideal for implementations on near-term quantum technologies, either quantum annealers or gate-based quantum computers. Such a low-dimensional compact representation is often stochastic and binary, in generative modeling~\cite{Bengio-Book}. We expect quantum devices to have a higher impact at processing this abstract representation, where the classically-tractable information has been already trimmed by the classical deep learning architecture. The lower layers of the network are classical components that effectively transforms samples from the quantum device to data points, and vice-versa. Hence, visible variables could be continuous variables, discrete variables, or other objects, effectively solving the encoding problem. (In Appendix~\ref{a:continuous} we argue why a direct implementation of stochastic continuous variables in hardware would be challenging even for the most trivial cases.) Finally, because the quantum device works on a low-dimensional binary representation of the data, we are also able to handle datasets whose dimensionality is much larger than it would be possible with state-of-the-art hardware.

The structure of the article is as follows: In Sec.~\ref{s:architectures} we describe some of the deep learning architectures that can be used in our framework. In Sec.~\ref{s:model} we formally define the quantum-assisted Helmholtz machine (QAHM) and derive the corresponding quantum-assisted wake-sleep learning algorithm. In Sec.~\ref{s:experiments} we describe some experimental results on the quantum-assisted generation of gray-scale handwritten digits of the MNIST dataset. In Sec.~\ref{s:summary} we present the conclusions and suggest future work. 

\section{Quantum-assisted architectures}~\label{s:architectures}

A deep generative model is based on a probability distribution $P(\bv) = \sum_{\bu} P(\bv|\bu) P(\bu)$, where ${\bv=\{v_1,\dotsc , v_N\}}$ are visible variables encoding the data and $\bu=\{u_1,\dotsc , u_M\}$ are unobserved or hidden variables that serve to capture non-trivial correlations by encoding high-level features. To perform inference and learning on this model, we have to sample from the posterior distribution $P(\bu|\bv)$, which is intractable in general. A standard approach to this problem consists of introducing a distribution $Q(\bu|\bv)$ to approximate the true posterior. When choosing the family of such distribution, one should consider functional forms that are both expressive and tractable. The learning algorithm is then in charge of adjusting $P(\bv)$ to model the data, and adjusting $Q(\bu|\bv)$ to approximate $P(\bu|\bv)$.

\begin{figure*}
\includegraphics[width=1.00\textwidth]{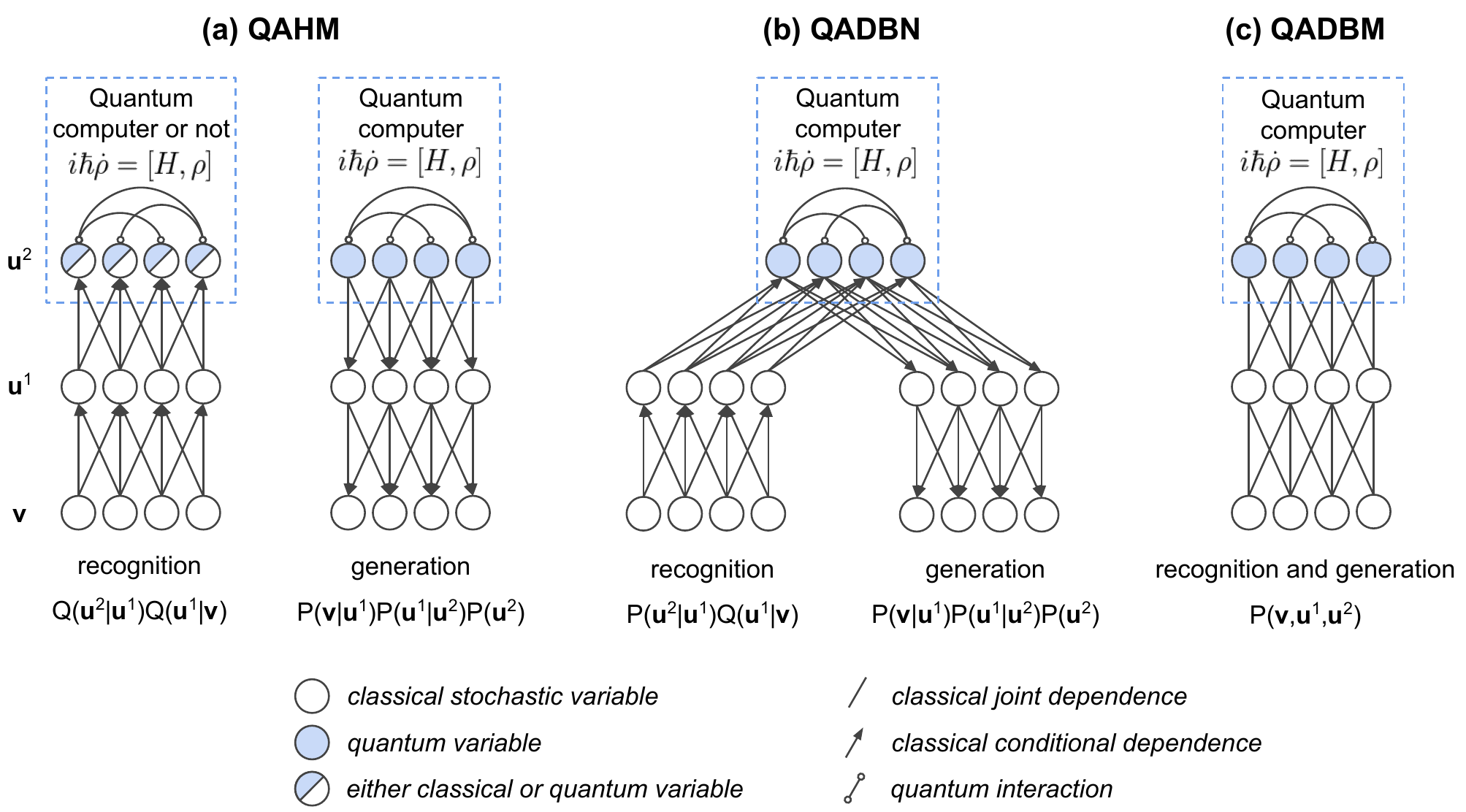}
\caption{Architectures for quantum-assisted machine learning (QAML). (a) quantum-assisted Helmholtz machine (QAHM); (b) quantum-assisted deep belief network (QADBN); (c) quantum-assisted deep Boltzmann machine (QADBM). We refer the reader to Sec.~\ref{s:architectures} for a brief description of the proposals pictured here.}
\label{f:qa_models}
\end{figure*}

We now consider some deep architectures that could work in synergy with quantum devices. In Fig.~\ref{f:qa_models}, generative models are represented as graphs of stochastic nodes where edges may be directed and undirected. We use the blue color for nodes that can be implemented on a quantum device, and we use an edge marked at both ends to indicate a quantum interaction. Fig.~\ref{f:qa_models}~(a) shows an instance of a Helmholtz machine~\cite{hinton1995wake, dayan1995helmholtz, bornschein2016bidirectional}, which consists of two networks: a {\it recognition network} to do approximate inference on hidden variables using information extracted from real data, and a {\it generator network} to generate artificial data. The recognition network implements the distribution $Q(\bu|\bv)$ and is used to perform bottom-up sampling starting from any visible vector $\bv$. This network may be entirely classical or quantum-assisted as discussed in Sec.~\ref{s:model}. The generator network, instead, implements the distribution $P(\bu,\bv)$ and is used to perform top-down sampling starting from the deepest hidden layer (e.g. $\bu^{2}$ in Fig.~\ref{f:qa_models}). The deepest hidden layer is modeled by quantum variables and quantum interactions. If the recognition and generator networks share the same quantum layer, we obtain the quantum-assisted version of a deep belief network~\cite{hinton2006fast, Bengio-Book} (QADBN; see Fig.~\ref{f:qa_models}~(b)). Deep belief networks usually implement a bipartite undirected graph in the deepest layer, but here we schematically show a more general structure with lateral connections that could be implemented in quantum hardware. Finally, if the recognition network is the exact inverse of the generator network, we obtain a quantum-assisted deep Boltzmann machine~\cite{ackley1985learning, salakhutdinov2009deep} (QADBM; see Fig.~\ref{f:qa_models}~(c)).

All three quantum-assisted architectures can be readily implemented and tested on available quantum computers. However, there is a practical caveat related to the fact that deep learning architectures require large datasets. For each data point, we need to perform recognition, and that requires both QADBN and QADBM to sample from a quantum device. This amount of work would be daunting for near-term quantum computers in the case of modern datasets. The more flexible framework of QAHM opens up the possibility of using a classical recognition network, sidestepping such limitation. We now discuss the details of the QAHM.

\section{Model definition and learning algorithm}\label{s:model}

Consider a dataset ${\mcS=\{\bv^1 ,\dotsc ,\bv^d\}}$ with empirical distribution $Q_\mcS(\bv)$. We seek a generative model ${P(\bv) = \sum_{\bu} P(\bu,\bv)}$, where ${P(\bu,\bv) = P(\bv|\bu) P_{QC}(\bu)}$. 

The prior distribution ${P_{QC}(\bu) = \langle\bu|\rho|\bu\rangle}$ describes samples obtained from a quantum device. For example, it could correspond to the diagonal elements of a quantum Gibbs distribution ${ \rho = e^{-\beta \mathcal{H}}/\mcZ }$, where $\mathcal{H}$ is the Hamiltonian implemented in quantum hardware and $\mcZ$ is the partition function. For instance, in the case of quantum annealing hardware, we could have
\be\label{e:ham}
\mathcal{H} = \sum_{i<j} J_{ij}\hat{Z}_i \hat{Z}_j + \sum_i h_i \hat{Z}_i +\Gamma\sum_i \hat{X}_i,
\ee
where $\hat{Z}_i$ and $\hat{X}_i$ denote Pauli matrices in the $z$ and $x$ direction, respectively, while $J_{ij}$, $h_i$, and $\Gamma$ are controllable parameters.

The conditional distribution $P(\bv|\bu)$ stochastically translates samples from the quantum computer into samples on the domain of the data. That is, $\bv$ could be a vector of continuous variables, binary variables, or other objects. This is a significant advantage over other quantum-assisted approaches where the visible variables are directly represented by qubits.

Ideally, an unsupervised learning algorithm would maximize the average log-likelihood of the data
\be\label{e:lik}
\mcL =\sum_\bv Q_\mcS(\bv)\ln P(\bv).
\ee
However, the training of a Helmholtz machine is based on the lower bound
\be\label{e:HM_lik}
\sum_\bv Q_\mcS(\bv) \ln P(\bv) \geq \sum_{\bv,\bu} Q_\mcS(\bv) Q(\bu |\bv)\ln\frac{P(\bu , \bv)}{Q(\bu | \bv)},
\ee
where $Q(\bu| \bv)$ is an auxiliary recognition network that approximates the intractable true posterior $P(\bu |\bv)$. Indeed, the name of the model comes from the minimization of the non-equilibrium Helmholtz free energy which is contained in the equation above~\cite{dayan1995helmholtz}. Our hybrid architecture uses a classical neural network for $Q$, sidestepping the need to sample from a quantum device for each data point and at each iteration of learning. This bottleneck is intrinsic in all the proposals we know up to date that treat quantum annealers as Boltzmann machines on the hidden layers of a neural network (e.g. see Ref.~\cite{potok2016study} for one recent such proposals). 

From now on, we focus on the case of quantum Gibbs distributions. The term ${\ln \langle\bu|\rho|\bu\rangle}$ arising from ${\ln P(\bu,\bv)}$ in Eq.~\eqref{e:HM_lik} is intractable due to the projection of the Gibbs distribution on the states $|\bu\rangle$. A bound for this term was derived in Ref.~\cite{Amin-arXiv-2016} using the \mbox{Golden-Thompson} inequality. Instead, we use a simpler bound based on Jensen's inequality (see Appendix~\ref{a:jensen} for a derivation)
\be\label{e:rho_ineq}
\ln\langle\bu |\rho|\bu\rangle \geq \langle\bu |\ln \rho|\bu . \rangle 
\ee 

Combining Eqs.~\eqref{e:HM_lik} and~\eqref{e:rho_ineq}, we get a tractable lower bound to maximize, i.e. the function
\be\label{e:G}
\mcG(\theta_G,\theta_{QC}) = \sum_{\bv,\bu} {Q_\mcS(\bv)}Q(\bu |\bv) \left[ \ln P(\bv | \bu) + \langle\bu|\ln\rho|\bu\rangle\right],
\ee
where $\theta_G$ and $\theta_{QC}$ denote the parameters of generator network $P(\bv|\bu)$ and quantum state $\rho$, respectively. In Eq.~\eqref{e:G} we neglected terms that do {\it not} depend on either $\theta_G$ or $\theta_{QC}$, as they vanish when computing the gradient of $\mcG$.

For a successful inference, the recognition network $Q(\bu|\bv)$ has to closely track the true posterior during learning. It is easy to see that the bound in Eq.~\eqref{e:HM_lik} is tight for ${Q(\bu|\bv) = P(\bu|\bv)}$. Unfortunately, the maximization of the lower bound in Eq.~\eqref{e:HM_lik} with respect to the parameters of the recognition network is often intractable. The wake-sleep algorithm~\cite{hinton1995wake} attempts to bring $Q(\bu|\bv)$ closer to the true posterior $P(\bu|\bv)$ by minimizing a more tractable notion of distance. Such distance is the Kullback-Leibler divergence 
\be\label{e:HM_D}
D_{KL}\left[P(\bu | \bv)\middle|| Q(\bu | \bv)\right] = \sum_{\bu}P(\bu | \bv)\ln \frac{P(\bu | \bv)}{Q(\bu | \bv)},
\ee
averaged over the marginal $P(\bv)$ to take into account the relevance of each configuration $\bv$. In other words, wake-sleep maximizes the function
\be\label{e:R}
\mcR(\theta_R) = \sum_{\bu,\bv}P(\bu , \bv)\ln {Q(\bu | \bv)},
\ee
where $\theta_R$ denotes, collectively, the parameters of the recognition network $Q(\bu|\bv)$. In Eq.~\eqref{e:R} we neglected terms that do not depend on $\theta_R$, as they vanish when computing the gradient of $\mcR$. 

The gradient ascent equations have structure ${\theta^{(t+1)} = \theta^{(t)} + \eta \nabla_\theta\mcF}$, where $\theta$ stands for the parameters being updated, $\eta$ is the learning rate, and $\mcF$ stands for either $\mcG$ or $\mcR$, accordingly. Since ${\ln\rho = -\beta \mathcal{H} - \ln\mcZ}$, for the parameters of the quantum distribution ${\theta_{QC} = (J_{ij},h_i)}$ we have
\BE \label{e:grad_J}
-\frac{1}{\beta}\frac{\partial \mcG}{\partial J_{ij}} &=& \langle u_i u_j\rangle_Q - \langle u_i u_j\rangle_\rho,
\EE
\BE \label{e:grad_h}
-\frac{1}{\beta}\frac{\partial \mcG}{\partial h_{i}} &=& \langle u_i \rangle_Q - \langle u_i \rangle_\rho,
\EE
where $\langle\,\rangle_Q$ and $\langle\,\rangle_\rho$ denote expectation values with respect to $Q(\bu|\bv)Q_\mcS(\bv)$ and ${P_{QC}(\bu) = \langle\bu|\rho|\bu\rangle}$, respectively. Here we have used the property ${\hat{Z}_i|u_i\rangle = u_i |u_i\rangle}$.

The generation and recognition networks can be written as deep learning architectures 
\be
P(\bv|\bu) = \sum_{\bu^1,\dotsc , \bu^L}P_0(\bv|\bu^1)P_1(\bu^1|\bu^2)\cdots P_L(\bu^{L}|\bu),
\ee
\be
Q(\bu|\bv) = \sum_{\bu^1,\dotsc , \bu^L}Q_L(\bu|\bu^L)\cdots Q_1(\bu^2|\bu^1) Q_0(\bu^1|\bv),
\ee
in terms of $L$ additional sets of hidden variables ${\bu^1,\dotsc ,\bu^L}$ that connect the variables ${\bv\equiv\bu^0}$ in the visible layer with ${\bu\equiv\bu^{L+1}}$ in the last hidden layer. More specifically, when using Bernoulli variables ${u_i^\ell \in \{-1,+1\}}$, we have
\BE \label{e:dist_p}
P_\ell(\bu^\ell|\bu^{\ell +1}) &=& \prod_i \pi(u_i^\ell | \bu^{\ell + 1};A^\ell, a^\ell),
\EE
\BE \label{e:dist_q}
Q_\ell(\bu^{\ell}|\bu^{\ell-1}) &=& \prod_i \pi(u_i^{\ell} |\bu^{\ell-1};B^{\ell}, b^{\ell}),
\EE
where 
\be\label{e:sigmoid}
\pi(u_i |\bu^\prime;C, c) = \left[1+ e^{-2 u_i\left( \sum_j C_{ij} u^\prime_j + c_i\right)}\right]^{-1}.
\ee

The gradients for the generative network are
\BE
\frac{\partial\mcG}{\partial A_{ij}^\ell} &=& \langle u_i^\ell u_j^{\ell +1} \rangle_Q - \langle u_i^\ell\rangle_P\langle u_j^{\ell+1}\rangle_Q,\\
\frac{\partial\mcG}{\partial a_{i}^\ell} &=& \langle u_i^\ell \rangle_Q - \langle u_i^\ell\rangle_P,
\EE
and similarly for the recognition network
\BE
\frac{\partial\mcR}{\partial B_{ij}^\ell} &=& \langle u_i^{\ell} u_j^{\ell-1} \rangle_P - \langle u_i^{\ell}\rangle_Q\langle u_j^{\ell-1}\rangle_P,\\
\frac{\partial\mcR}{\partial b_{i}^\ell} &=& \langle u_i^{\ell} \rangle_P - \langle u_i^{\ell}\rangle_Q.
\EE

We now discuss some alternatives and improvements that can be found in the literature of deep generative models. A generalization of the wake-sleep algorithm, called reweighted wake-sleep, was introduced in Ref.~\cite{bornschein2014reweighted}. The authors used $Q$ as a proposal distribution for importance sampling of $P$, and obtained a better gradient estimator by reducing bias and variance. Another approach was introduced in Ref.~\cite{salakhutdinov2010efficient} in the context of deep Boltzmann machines. Samples from $Q$ were used as starting points for a set of mean-field equations; the mean-field solutions provided a closer approximation to the expectation values required for training. Finally, there exists a contrastive version of the wake-sleep algorithm that was introduced in Ref.~\cite{hinton2006fast} to train deep belief networks with undirected edges. In contrastive wake-sleep, samples from $Q$ are used to seed a Gibbs sampler for the deepest layer of $P$, aiding thermalization. 

All the improved techniques discussed above require full knowledge of the parameters. This may not be available in noisy quantum annealers or quantum devices without error correction. Nevertheless, we now show how the vanilla wake-sleep algorithm can be used to train Helmholtz machines assisted by noisy quantum annealers. Advantages, challenges and potential generalizations are discussed in Sec.~\ref{s:summary}.

\section{Experiments}\label{s:experiments}

We demonstrate the QAHM framework using a \mbox{D-Wave 2000Q} quantum annealer hosted by the NASA Ames Research Center. The annealer implements a noisy version of the programmed Hamiltonian in Eq.~\eqref{e:ham} defined on a sparse graph of qubit interactions. In particular, the device is designed to exploits quantum tunneling to sample low energy states at transverse field $\Gamma\approx 0$. However, non-trivial non-equilibrium effects may make samples deviate from the corresponding classical Gibbs distribution. This scenario requires some engineering of the QAHM framework as well as additional actions besides those outlined in Sec.~\ref{s:model}. We would like to stress that the algorithm can be carried out on other quantum annealing architectures~\cite{lechner2015quantum,PerdomoOrtiz2017a}, and on more general gate-based quantum computers. Implementations in these architectures may require further, or fewer, engineering steps, and could allow more general quantum distributions.

Following the work in Ref.~\cite{Benedetti2017}, we use a gray-box model for the quantum annealer so that we can update its parameters without the need to estimate deviations from the Gibbs distribution. This approach relies on the assumption that, despite the deviations, the estimated gradients have a positive projection in the direction of the true gradient. Because of a varying unknown inverse temperature $\beta$, the learning rate at which parameters are updated varies too. This should not pose a problem as long as we schedule the learning rate to decrease, which is a general condition for convergence of stochastic approximation algorithms of \mbox{Robbins-Monro} type~\cite{younes1999convergence}. 

Now, we would like to implement a fully connected prior distribution $P_{QC}(\bu)$ over hidden variables in the deepest layer. This connectivity in not available in hardware, so we map each variable to a subgraph of physical qubits. This way, the additional physical interactions between qubits can effectively encode long-range interactions. This expansion needs not be globally optimal, and can be found efficiently using heuristic techniques. The new dynamics are described by the programmed Hamiltonian
\be \label{e:core}
\widetilde{\mathcal{H}} = -\frac{1}{2} \sum_{i,j=1}^N \sum_{k,l=1}^{Q_{i},Q_{j}} J_{ij}^{(kl)} \hat{Z}_{i}^{(k)} \hat{Z}_{j}^{(l)} - \sum_{i=1}^N \sum_{k=1}^{Q_{i}} h_{i}^{(k)} \hat{Z}_{i}^{(k)}.
\ee
Here $N$ is the number of hidden variables in the deepest layer, which equals the number of subgraphs realized in hardware, $Q_i$ is the number of qubits in subgraph $i$, $\hat{Z}_{i}^{(k)}$ is the Pauli matrix in the $z$-direction for qubit $k$ of subgraph $i$, $h_{i}^{(k)}$ is the local field for qubit $k$ of subgraph $i$, and $J_{ij}^{(kl)}$ is the coupling between qubit $k$ of subgraph $i$ and qubit $l$ of subgraph $j$. Note that the couplings serve to model both the consistency within subgraphs, when $i=j$, and the correlation among subgraphs, when $i\neq j$. A factor of $1/2$ is required to avoid double counting. The gradients required to learn these parameters are similar to those in Eqs.~\eqref{e:grad_J} and~\eqref{e:grad_h}, and can also be found in Ref.~\cite{Benedetti2017}.

\begin{figure*}
\includegraphics[width=1\textwidth]{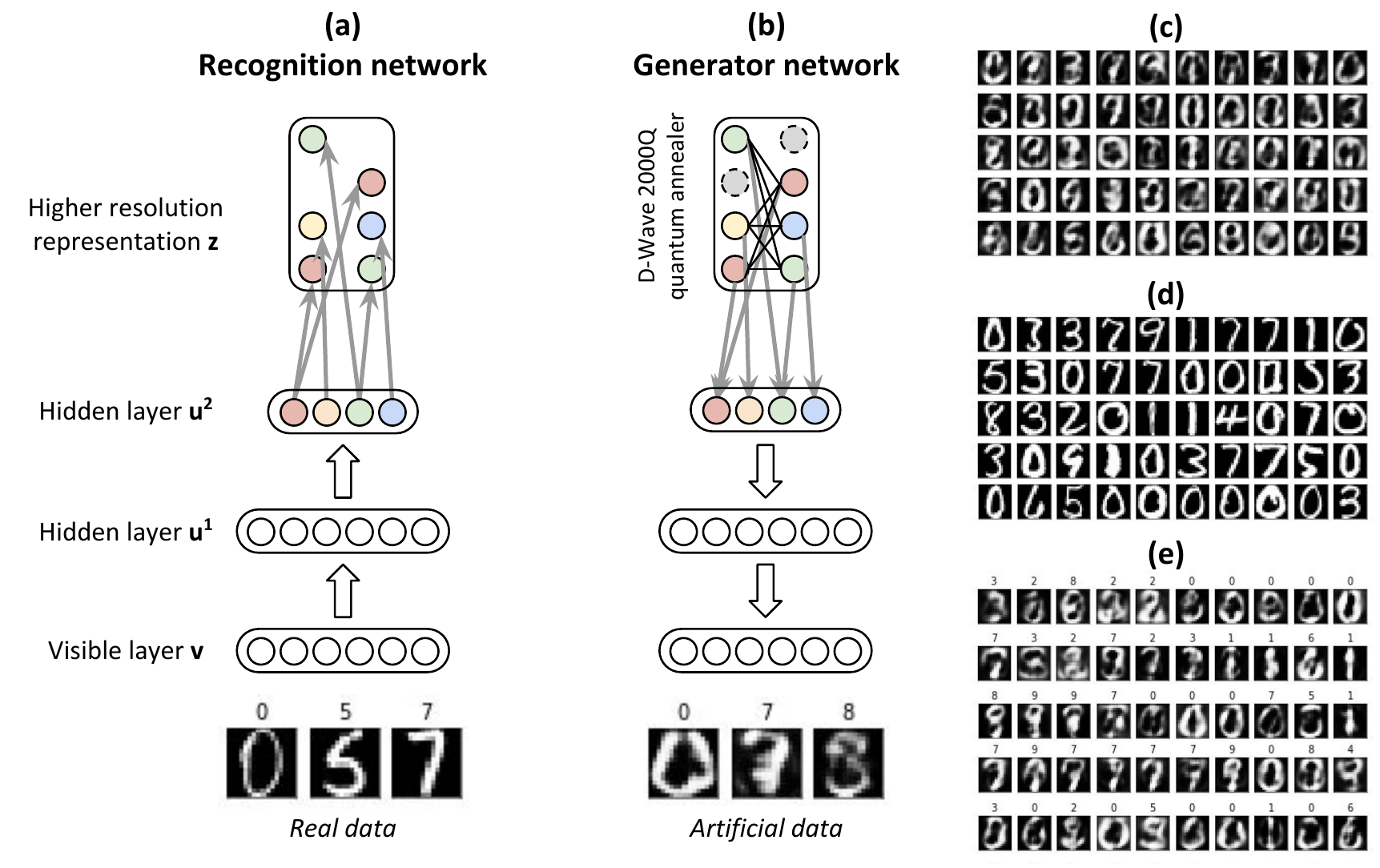}
\caption{Scheme for the experimental implementation of the QAHM on the \mbox{D-Wave 2000Q} quantum annealer for the sub-sampled MNIST dataset. The visible layer consists of 256 continuous variables $\bv$ that encode the gray-scale pixels of \mbox{16 $\times$ 16} images, and 10 binary variables that encode the class. There are two hidden layers, $\bu^1$ and $\bu^2$, with 120 and 60 hidden binary variables, respectively. The variables $\bu^2$ in the second hidden layer are effectively connected all-to-all through an embedding into 1644 qubits, $\bz$, of the quantum annealer (see Ref.~\cite{Benedetti2017} for details). The recognition network~(a) is entirely classical to avoid calling the quantum device for each point in the dataset (7291 images from the sub-sampled version of the MNIST handwritten dataset used here). The generator network~(b) samples the deepest layer from the quantum annealer. The necessary correspondence between recognition and generator networks is enforced by two deterministic mappings, here represented by gray-colored edges. Panel~(c) shows artificial images obtained from the generator network after training. Panel~(d) shows the images in the training set that are closest in Euclidean distance to the artificial ones in panel~(c). Note that artificial images are not merely copies of the training images. Panel~(e) shows additional artificial images along with their most probable class according to the model. Visually, the quantum-assisted model seems to correlate class and pixels most of the time.}
\label{f:dw_qahm}
\end{figure*}

The model is also equipped with two deterministic functions that map samples back and forth between the two spaces (i.e. logical and qubit spaces). We use the following {\it replica} and {\it majority vote} mappings
\be \label{e:replica}
z_{i}^{(k)} = f(\bu, i) = u_i \quad (\text{for } k=1,\dots,Q_{i}),
\ee
\be \label{e:majority}
u_{i} = g(\bz, i) = \sign \left( \sum_{k=1}^{Q_{i}} z_i^{(k)} \right).
\ee 
These mappings can be thought of as non-trainable edges in the recognition and generator networks, respectively. To see why, consider a QAHM with one visible $\bv$ and two hidden layers $\bu^1$ and $\bu^2$, like the one shown in Figs.~\ref{f:dw_qahm}~(a) and~\ref{f:dw_qahm}~(b). In the recognition network, the hidden variables $\bu^2$ get replicated into higher-dimensional vectors $\bz$ (replicas are shown with the same color). We can easily sample from the recognition network using a bottom-up pass that does not involve the quantum device. In the generator network instead, the quantum device is used to sample $\bz$ from a Gibbs-like distribution. Samples are mapped back to the hidden variables $\bu^2$ using the majority vote over subgraphs (subgraphs are shown with the same color). Then, a top-down pass is used to sample the visible variables $\bv$. Hence, every directed and undirected edge in Fig.~\ref{f:dw_qahm} can be trained, except for the gray-colored directed edges corresponding to the fixed mappings in Eqs.~\eqref{e:replica} and~\eqref{e:majority}. In future work, we will consider extending the model by including a quantum device in the deepest layer of the recognition network. This will require to sample from the device conditionally on each data point.

\begin{figure*}
\begin{minipage}{\linewidth}
\begin{algorithm}[H]
\caption{Wake-sleep algorithm for quantum-assisted Helmholtz machines on quantum annealers}
\label{a:ws_dw}
\begin{algorithmic}
\STATE use an heuristic to embed in hardware a fully connected graph corresponding to the deepest hidden layer
\STATE define mappings $f(\bu, i)$ and $g(\bz, i)$ from hidden variables to qubits and back
\FOR{number of training epochs}
  \STATE sample $(\bv^d,\bu^d,\bz^d)$ where $ (\bv^d, \bu^d) \sim Q(\bu|\bv)Q_{\mcS}(\bv)$ and $z^{d}_i=f(\bu^d,i)$
  \STATE sample $(\bv^k,\bu^k,\bz^k)$ where $\bz^k \sim \langle\bz|\rho|\bz\rangle$, $u^{k}_i = g(\bz^k,i)$ and $\bv^k \sim P(\bv|\bu^k)$    
  \STATE estimate $\nabla_{\theta}\mcG$ and $\nabla_{\theta}\mcR$ from samples
  \STATE update $\theta^{(t+1)}_\mcG = \theta^{(t)}_\mcG + \eta \nabla_\theta\mcG$
  \STATE update $\theta^{(t+1)}_\mcR = \theta^{(t)}_\mcR + \eta \nabla_\theta\mcR$
  \STATE decrease $\eta$
\ENDFOR
\end{algorithmic}
\end{algorithm}
\end{minipage}
\end{figure*}

Now, because we don't have complete knowledge of the parameters implemented by the annealer, we cannot use techniques such as importance sampling that have been used to improve the wake-sleep algorithm and obtain state-of-the-art results (see Section~\ref{s:model} for a brief summary). We shall stress that this limitation is peculiar of our case-study and may not be present in other quantum hardware (e.g. error-corrected quantum computers). Improved and faster learning can also be obtained by initializing the approximate posterior $Q(\bu|\bv)$ close to true posterior $P(\bu|\bv)$ when $\bv$ is sampled from the dataset. This initialization, also called {\it pre-training}, is often carried out by stacking layers of restricted Boltzmann machines and training them greedily with some fast approximate algorithm~\cite{hinton2006fast, salakhutdinov2010efficient}. In principle, we could use pre-training to initialize all the trainable directed edges of our model (see Fig.~\ref{f:dw_qahm}). The procedure would trivially extend to the undirected edges in the generator network because the pre-trained recognition network would effectively provide a fully-observed dataset for computing the gradients in Eqs.~\eqref{e:grad_J} and~\eqref{e:grad_h}. We decided not to carry out pre-training in our small scale experiment as it could initialize the model to an almost-optimal configuration, hence hiding any contribution of the quantum device. For the reasons outlined above, we acknowledge that our vanilla wake-sleep algorithm may be slow and sub-optimal (this is further discussed in Section~\ref{s:summary}). The wake-sleep algorithm for Helmholtz machines on quantum annealers is summarized in Algorithm~\ref{a:ws_dw}.

We tested our ideas on a sub-sampled version of the MNIST handwritten digits dataset~\cite{subMNIST}. Our training set consists of $7291$ images of $16\times 16$ gray-scale pixels, and a categorical variable indicating the corresponding digit. First, we rescaled pixels to take real-values in $[-1,+1]$. Second, we used a one-hot encoding for the class (i.e. $c^d_i=-1$ for $i\neq j$, $c^d_j=+1$ where $j$ indexes the class for image $d$) obtaining $10$ binary variables. The visible layer was connected to a first hidden layer of $120$ binary variables which, in turn, was connected to a second hidden layer of $60$ binary variables. We used \mbox{D-Wave} heuristics~\cite{cai2014practical} to embed a fully connected graph of $60$ variables in the \mbox{D-Wave 2000Q}. This resulted in a graph of $1644$ qubits in total, where the largest subgraph had $43$ qubits and the smallest subgraph had $18$ qubits. The maps in Eqs.~\eqref{e:replica} and~\eqref{e:majority} were set up accordingly. Figure~\ref{f:dw_qahm} shows the final model composed of two networks and a quantum annealer implementing a prior over the second hidden layer, $\bu^2$. It can be easily seen that the final model is an engineered version of the model in Fig.~\ref{f:qa_models}~(a). To implement the continuous variables, $\bv$, we used a deterministic layer of hyperbolic tangent non-linearities, which is compatible with our rescaling in the interval $[-1,1]$. Alternatively, one can use stochastic Gaussian variables and a different, compatible, rescaling.

We ran the vanilla wake-sleep algorithm for $500$ epochs with a learning rate of $0.005$ for all the gradient updates. Subsequently, we trained for other $500$ epochs by linearly decreasing the learning rate down to $0.0005$. At each training iteration, we inferred hidden configurations from the recognition network for all the data points in the training set, and sampled $1000$ artificial points from the generator network. These two sets are used to compute gradients as in Algorithm~\ref{a:ws_dw}. Quantum annealing hyperparameters such as annealing time, programming thermalization and readout thermalization were set to their corresponding minimum values in order to obtain samples as fast as possible. Of particular importance, the annealing time determines how fast the quantum computing environment evolves towards the programmed Hamiltonian in Eq.~\eqref{e:core}. The use of the minimum annealing time is a well established practice due to extensive benchmarking by the combinatorial optimization community. We are not aware of similar systematic studies in the context of sampling, although we expect annealing time to have a significant impact on the form of the distribution. Because the gray-box model considered here does not require knowledge of the exact form of the distribution, we chose the minimum annealing time of $5\mu s$. 

Figure~\ref{f:dw_qahm}~(c) shows samples from the generator network after training. For each of those, Fig.~\ref{f:dw_qahm}~(d) shows the image in the training set that is closest in Euclidean distance. We can see that the artificial data generated by the model is \textit{not} merely a copy of the training set. The generated data presents variations and, in some cases, novelty, reflecting the generalization capabilities of the model. Although these preliminary results cannot compete with state-of-the-art ML, the generated data often resemble digits written by humans. Indeed, the problem of generating blurry artificial images affects other approaches as well; only the recent development of generative adversarial networks~\cite{goodfellow2014generative} led to much sharper artificial images. 

Finally, Fig.~\ref{f:dw_qahm}~(e) shows some artificial samples along with their most probable class according to the model. Visually, the model seems to correlate class and pixels most of the time. The process can be easily generalized to perform classification, where test images are provided through the recognition network and the most likely class is inferred through the generator network.

\section{Conclusions and future work}\label{s:summary}

Despite significant effort in quantum-assisted machine learning (QAML), there has been a disconnect between most algorithmic proposals, the needs of machine learning (ML) practitioners, and the capabilities of near-term quantum devices. Inspired by the challenges and guidelines exposed in Ref.~\cite{PerdomoOrtiz2017}, we implemented a hybrid classical-quantum architecture for unsupervised learning. We demonstrated how currently available quantum devices can be used in real-world modeling applications on datasets with higher dimensionality than apparently possible, and on variables which are not binary, e.g. modeling of gray-scale handwritten digits of $16\times 16$ pixels. In our case study, we used a noisy quantum annealer to learn an implicit prior distribution for the latent variables of a deep generative model. 

Here, we summarize some of the advantages and challenges with the current implementation of the quantum-assisted Helmholtz machine (QAHM), and we propose some generalizations for future work.\\

\noindent{\textit{Advantages of the QAHM framework}:}
\begin{itemize}
\item A classical recognition network is used to perform approximate inference. There is no need to sample from a quantum device for each data point and for each learning iteration.
\item The quantum device is employed in the deepest layers of a generator network. The lowest layers stochastically transform the information from qubits to data vectors, and back. Data vectors can be discrete, continuous, or of a more general type.
\item The quantum device models an abstract representation whose dimensionality is expected to be much smaller than that of the raw data. This enables the handling of datasets of relevant size, a significant step towards real-world applications.
\end{itemize}

\noindent{\textit{Challenges and why our experiments are sub-optimal:} }

\begin{itemize}
\item The sleep phase of the wake-sleep algorithm optimizes the wrong cost function~\cite{hinton1995wake}. Solutions found in the literature~\cite{bornschein2016bidirectional, salakhutdinov2009deep} require full knowledge of the model's parameters which is not available under the gray-box approach employed here.
\item The recognition network has to be expressive enough to closely track the true posterior. As pointed out in the original work on Helmholtz machines~\cite{hinton1995wake}, factorized distributions are not able to model complex posteriors because of non-trivial effects such as {\it explaining away}. Studies shown that better likelihoods are obtained when the recognition network is equipped with more complex hidden layers (e.g. autoregressive or NADE)~\cite{bornschein2016bidirectional}. However, we expect the problem to be much more dramatic when using quantum distributions in the generative network as done here. This may require the introduction of a quantum distribution in the recognition network as well, hence losing one of the advantages listed above.
\end{itemize}

\noindent{\textit{Some potential generalizations:}} 
\begin{itemize}
\item The deterministic mappings in Eqs.~\eqref{e:replica} and~\eqref{e:majority}, used here to translate information from and to quantum hardware, can be relaxed into trainable functions. In this case, variables $\bz$ in the recognition network and $\bu^2$ in the generator network become stochastic Bernoulli variables. Indeed, the expected value of a Bernoulli variable ${u_i \in \{-1, +1\}}$, conditioned on the configuration $\bu^\prime$ of the previous layer, is described by the hyperbolic tangent function ${\fE \left[ u_i | \bu^\prime \right] = \tanh( c_i + \sum_{j} C_{ij} u^\prime_j)}$. When ${C_{ij}\gg 1}$ and $c_i=0$, this function implements a majority vote of the variables in the previous layer. The replica function can be thought of as a majority vote over a single qubit in the previous layer. Hence, by allowing all parameters $c_i$ and $C_{ij}$ to be learned, one obtains a generalized version of the quantum-assisted wake-sleep algorithm introduced here. While this generalization requires fitting additional parameters, it has the potential to discover better embeddings than those found via heuristics.
\item The general QAHM framework allows to use quantum devices in both the recognition and the generator networks (see Fig.~\ref{f:qa_models}~(a)). The motivation for using the quantum device only in the generator network is to bypass the issue of making calls to the quantum device for every point in the dataset. It is an open question whether using the quantum device in the recognition network can significantly enhance the quality of the model.
\end{itemize}

Although the results of the current implementation on quantum annealers do not compete with state-of-the-art computer vision systems, we hope this flexible QAHM framework will motivate researchers to develop novel hybrid quantum-classical approaches, with the intention to use near-term quantum computers for intractable tasks such as unsupervised learning and sampling.

\section*{Acknowledgements}

The work of A.P-O., J-R-G, and M.B.~ was supported in part by the AFRL Information Directorate under grant \mbox{F4HBKC4162G001}, the Office of the Director of National Intelligence (ODNI), the Intelligence Advanced Research Projects Activity \mbox{(IARPA)}, via IAA 145483, and the U.S. Army TARDEC under the ``Quantum-assisted Machine Learning for Mobility Studies" project. The views and conclusions contained herein are those of the authors and should not be interpreted as necessarily representing the official policies or endorsements, either expressed or implied, of ODNI, IARPA, AFRL, U.S. Army TARDEC or the U.S. Government. The U.S. Government is authorized to reproduce and distribute reprints for Governmental purpose notwithstanding any copyright annotation thereon. M.B. was partially supported by the UK Engineering and Physical Sciences Research Council \mbox{(EPSRC)} and by Cambridge Quantum Computing Limited \mbox{(CQCL)}. 

\appendix
\section{Approximating continuous stochastic variables in quantum annealers}\label{a:continuous}

Here we show how naive approaches to encoding continuous variables in quantum annealers are likely to fail. Consider the task of approximating a simple univariate Gaussian probability. If we were able to do that, we could control its mean $\mu$ and variance $\sigma^2$, and sample accordingly. While this is a trivial task in classical computers, it serves as an example to show the challenge of implementing continuous variables in quantum annealers. One way to approach the problem is to approximate the stochastic continuous variable $x$ with the weighted sum of a large number of qubits, i.e. ${x=\sum_i w_i s_i}$ where $w_i$ are programmable weights in the annealer. Notice that $n$-ary expansions commonly used in classical computers are just special cases of this weighted sum where weights increase or decrease exponentially with the precision (i.e. number of qubits used for the encoding). This would not be practical for state-of-the-art devices as it requires high-precision parameters that are not available because of noise, bias, and finite control precision. A more general weighted-sum encoding may introduce degeneracy, but this is not a problem in the machine learning setting considered here as long as the results approximate the desired continuous probability distribution. Moreover, in the machine learning setting we could learn all the parameters, including the weights $w_i$. 

Now, consider approximating the Gaussian probability over $x$ in the annealer. We define an energy function encoding the eigenvalues of the Hamiltonian in Eq.~\eqref{e:ham} with zero transverse field
\be
\begin{split}
E(\bs) &= \frac{1}{2 \sigma^2 } \Big ( \sum_i w_i s_i - \mu \Big)^2 \\
&= \frac{1}{2 \sigma^2} \Big( \sum_{i \ne j} w_i w_j s_i s_j + \sum_{i} w_i^2 + \mu^2 - 2 \mu \sum_i w_i s_i \Big) \\
&= \sum_{i \ne j} J_{ij} s_i s_j + \sum_i h_i s_i + C 
\end{split}
\ee
where ${J_{ij} = { w_i w_j}/{2 \sigma^2}}$ are couplings, ${h_i=-{\mu w_i}/{\sigma^2}}$ are local fields, and we collected the constant terms in $C$. The result is a fully connected graph that must be natively implemented in hardware. That is, if we want $N$-bits of precision, we are required to have an $N$-clique in the hardware interaction graph. To see why, assume one of the interactions is not available in hardware, that is $J_{ij}=0$. From the definition of $J_{ij}$ above, we see that either $w_i=0$ or $w_j=0$. Take $w_i=0$ and notice that $J_{ik}=0$ for each $k$, or in words, qubit $i$ is disconnected from the interaction graph. Then, qubit $i$ is useless for the purpose of approximating the desired continuous variable. As an example, the chimera interaction graph used in the \mbox{D-Wave 2000Q} has a largest clique of size $2$. Hence, the best naive encoding has $2$ bits of precision, and they are clearly not enough to approximate and have control over any desired Gaussian distribution.

While in this specific instance a simple solution is possible through the central-limit theorem, and more elaborated approaches may also be possible, this discussion suggests that the implementation of stochastic continuous variables may be challenging in more general setups that go beyond the univariate Gaussian case.

\section{Derivation of the bound for quantum Gibbs distributions}\label{a:jensen}

We require a tractable bound for ${\ln\langle\bu |\rho|\bu\rangle}$ in order to train the QAHM when a quantum Gibbs distributions is used in the generator network. First, write the density matrix in terms of eigenvectors $|i\rangle$ and eigenvalues $E_i$ of the Hamiltonian
\be
\rho = \sum_i \frac{e^{-E_i}}{\mcZ} |i\rangle \langle i| ,
\ee
where $\mcZ=\sum_i e^{-E_i}$ is the normalization constant. Then, plug this expansion into the intractable expression and use Jensen's inequality
\be\label{e:jensen}
\begin{split}
\ln\langle\bu |\rho|\bu\rangle &= \ln\langle\bu | \sum_i \frac{e^{-E_i}}{\mcZ} |i\rangle \langle i | \bu\rangle \\
&= \ln \sum_i |\langle i|\bu\rangle|^2 \frac{e^{-E_i}}{\mcZ} \\
&\geq \sum_i |\langle i|\bu\rangle|^2 \ln \frac{e^{-E_i}}{\mcZ} \\
&= \langle\bu | \sum_i \ln \frac{e^{-E_i}}{\mcZ} |i\rangle \langle i | \bu\rangle \\
&= \langle\bu |\ln \rho|\bu\rangle ,
\end{split}
\ee 
where $|\langle i|\bu\rangle|^2$ are probabilities and sum up to 1.


%

\end{document}